\documentstyle[aps,preprint]{revtex}

\begin{document}

%@twocolumnfalse\endcsname

\vspace{2cm}
\title{Universality, Gravity, the enigmatic $\Lambda$ and Beyond}
\author{Naresh Dadhich \footnote{email: nkd@iucaa.ernet.in}} 
\address {Inter-University Centre for Astronomy \& Astrophysics,
Post Bag 4, Pune~411~007, India}

\maketitle

\begin{abstract}
In this essay, I wish to share a novel perspective which envisions 
universalization as a guide from the classical world to relativistic and 
quantum world. It is the incorporation of zero mass particle in mechanics 
which leads to special relativity while its interaction with a universal 
field shared by all particles leads to general relativity. We also give a 
very simple classical argument to show that why the universal force has to 
be attractive. We try to envisage what sort of directions does this principle 
of universality point to for the world beyond general relativity?

\end{abstract}
\vskip2pc 

%\pacs{ 98.80.Cq, 98.70.Vc}

\newpage

 In the classical Newtonian World, we have the absolute $3$-space 
and the absolute $1$-time with space having the Euclidean flat metric. Then 
there are the three laws of motion governing motion of particles. The laws 
however apply only to massive particles. That is massless particles cannot be 
accommodated in the Newtonian mechanics. If such particles exist in nature, 
they would ask for new mechanics. The characteristic of such a particle is 
that it can't exist at rest in any frame. It should be moving relative to all 
observers and its speed must hence be the limiting speed for all. It must be 
the same for all observers. Existence of zero mass particles gives us a 
universal constant speed which needs to be accommodated in mechanics. \\

 On the other hand we also know in the Newtonian framework that light always 
moves in a straight line in all directions. The straight line motion is 
indicative of constant speed. Not 
quite, there could occur accelerated motion in straight line for force acting 
along the line of motion. Since light moves straight in all directions, this 
possibility is also ruled out because there cannot exist a force acting in 
all directions. We are thus forced to conclude that either 
Newton's laws of motion do not apply to light or its speed must be constant. 
This constant should be the same for all observers because light moves in a 
straight line for all observers. That is, light's speed must also be a 
universal constant. \\

 We know that zero mass particles do indeed exist in nature and they are 
nothing but the particles of light, photons. Though they are strictly 
speaking quantum objects but they do have classical menifestation as 
electromagnetic wave. They could however be used to universalize 
mechanics;i.e. it must apply to both massive as well as massless particles. 
That then asks for a new mechanics which is special relativity (SR). \\

 The most fundamental universal entities we know are space and time and 
a universal statement is always expressed through geometry. For instance, 
motion of free particles is a straight line is a universal property which is 
expressed as a geometric statement. From this we wish to propound a general 
guiding principle, {\it any universal physical property/force must be 
expressible as a property of universal entities, space and time, through a 
geometric statement}.\\

 Thus the existence of universal speed, which we denote by $c$, must become a 
property of space and time 
and that is what precisely Minkowski did when he incorporated invariance of 
light's speed in the geometry of $4$-dimensional spacetime manifold. This 
bound space and time into one whole, spacetime. \\

 What we have done is that in conformity with universality we have asked for 
incorporation of massless particles in the Newtonian mechanics, which was not 
allowed. Taking into account the characteristic property of motion of massless 
particles we are forced to enlarge the Newtonian framework to Minkowski 
framework which admitted both massive and massless particles. In our further 
exploration, this would be the guiding finger;i.e. {\it ask a question which 
is not admitted in the existing framework, then enlarge the framework as 
indicated/suggested by the question itself such that the question is answered.}
\\

 Let us consider that there exists a universal force which is shared by all 
particles. It must also act on 
massless particles which propagate with universal constant speed. Then how do 
we make it feel the universal force, and feel it must. We thus face the 
contradiction that massless particle must feel the force but its speed must 
not change. This is not possible in the existing framework. How do we enlarge 
the framework? Two suggestions come forward from our guiding principle, one 
since the force is universal it must be expressible as a property of spacetime
 and two, what do we really want massless particle to do? What we want the 
particle to do is that when it is skirting the source of the universal force 
it should acknowledge its presence by bending rather than going 
straight. To illustrate this point, let 
us consider a piece of wood floating in a river. It floats freely and bends 
as the river bends. No force really acts on it to bend its course, it simply 
follows the flow of the river. This suggests that bend the {\it river} in 
which massless particle propagates freely. Thus we arrive at the profound 
insight into the nature of the universal force that it must bend/curve space - 
rather spacetime in which massive as well as massless particles propagate 
freely. That is the envisaged universal force can only be described by 
curvature of spacetime and 
in no other way. \\

 Thus universality of the force does not let spacetime remain inert 
background but instead impregnates it with dynamics, a unique and 
distinguishing property. It then ceases to be an external force and its 
dynamics has to be fully governed by the spacetime curvature. This 
automatically incorporates the first suggestion mentioned above that 
universal force should become a property of spacetime. The equation of motion 
for the force does not have to be prescribed but must rather follow from the 
spacetime curvature all by itself. The curvature of spacetime is given by the 
Riemann curvature 
tensor for the metric $g_{ab}$ and it satisfies the Bianchi 
differential identity. The contraction of 
which yields the second rank symmetric tensor, constructed from the Ricci 
tensor, having vanishing divergence. That is, 
\begin{equation} \label{div}
\nabla_{b}G^{ab} = 0 
\end{equation}
where 
\begin{equation}
G_{ab} = R_{ab} - \frac{1}{2}Rg_{ab}\,.
\end{equation}
The above equation implies 
\begin{equation} \label{gr}
G_{ab} = -\kappa T_{ab} - \Lambda g_{ab}
\end{equation}
with 
\begin{equation}
\nabla_{b}T^{ab} = 0
\end{equation}
where $T_{ab}$ is a symmetric tensor, and $\kappa$ and $\Lambda$ are 
constants. On the left we have a differential expression involving the second 
derivative and square of the first derivative of the metric which now acts 
as a potential for the universal force. Thus on the right should be the 
source of the force. What should be the source for a universal force? 
Something which is shared by all particles - mass/energy. That identifies 
$T_{ab}$ with the energy momentum tensor of matter distribution and vanishing 
of its divergence ensures conservation of energy and momentum. \\

 We however know that in the Newtonian theory mass is the source for gravity 
and hence the above equation should agree with the Newtonian equation in the 
first approximation. This determines $\kappa = 8\pi G/c^2$ in terms of the 
Newtonian constant of gravitation, and the new 
constant $\Lambda$ should be small so as to have no effect over the stellar 
scale where the theory has been tested observationally. We have thus obtained 
the Einstein equation for gravitation - general 
relativity (GR) \cite{n1,n2}. It is remarkable to note that the universal 
force can be nothing other than gravity. Thus gravity  is the unique 
universal force. \\ 
    
 Note that the constant $\Lambda$ enters into the equation naturally and is 
in fact at 
the same footing as the energy momentum tensor. It makes a very important 
statement that even when space is free of all non-gravitational matter/energy 
distribution, empty space has non-trivial dynamics. The above equation refers 
to spacetime in entirety which means whatever can be done in or to it is 
included in this equation. It is well known that vacuum can suffer quantum 
fluctuations which produce stress energy tensor preciely of the form 
$\Lambda g_{ab}$ and it must be 
included in the equation. Had Einstein followed this chain of arguments, he 
would have anticipated 
gravitational effect of quantum fluctuations of vacuum and would have made a 
profound prediction rather than a profound blunder. \\

 Classically a constant scalar field has no dynamics while in GR it again 
generates 
precisely the stress tensor of the type $\Lambda g_{ab}$ which of course has 
non trivial dynamics. Thus $\Lambda$ is 
the potential to which vacuum has been raised.
\footnote{ In the Schwarzschild field, we have the Newtonian potential, 
$\phi = 
-M/r$, note that a non-zero constant, which is classically inert, 
cannot be added to it.
The Einstein equation determines the potential of an isolated body absolutely 
with its zero fixed at infinity and nowhere else \cite{n3}. That is for the 
local situation and here we have constant potential in a global setting.} 
 That means when $\Lambda = 0$, 
the vacuum is at absolute zero potential. Since vacuum can suffer quantum 
fluctuations, it must have micro-structure which can fluctuate. So long as 
spacetime has micro-structure, which is necessarily required for all quantum 
phenomena in free space, the constant $\Lambda$ cannot in general be made to 
zero without introducing some new feature like supersymmetry. Even then, I 
would wonder whether zero $\Lambda$ vacuum could be stable? The 
big open question is what value should it have? It is in fact a new constant 
of the Einsteinian gravity which needs to be understood. \\

 Another aspect of the equation is that it is valid in all dimensions, 
$n\ge2$, for which Riemann curvature can be defined. It is so because gravity 
is universal in the most general sense and hence it cannot by hand be 
restricted to a given dimension. All by itself gravity is thus inherently a 
higher dimensional interaction. One can restrict $T_{ab}$ on the right to 
a given space dimension by confining matter field on it while $\Lambda g_{ab}$ 
refers to dynamics of vacuum which cannot like matter field be confined. 
Gravity is described by curvature of spacetime and to realise its full 
dynamics the minimum number of dimensions required are $4$. Gravitational 
charge is matter/energy which is always positive while gravitational field 
energy is negative. It is by taking the two together there can occur charge 
neutrality of the gravitational systems. Since the negative charge is spread 
over the entire space, the total charge will vanish only when you integrate 
over the whole space. That is in the local neighbourhood there will be net 
positive charge (energy distribution) which may be confined to $3$-space, yet 
gravitational field will propagate off into the higher dimension. Though it 
would not be able to penetrate deep into the higher dimension because 
globally there exists charge neutrality. If the matter fields are confined to 
$n$-space, the field will leak into the $(n+1)$th dimension but will not 
propagate deep enough. That is the massless graviton will have ground state 
and hence will remain confined to $n$-space. This is exactly what is required 
to happen for the brane world gravity \cite{add,rs}. \\ 

 It is remarkable that we have here motivated the higher dimensional and the 
brane world nature of the gravitational field purely from classical 
standpoint. If the matter is confined to 3-brane, the bulk 
spacetime would be 
$5$-dimensional with $\Lambda$ being its dynamical support. In particular we 
can have the Randall - Sundrum brane world model \cite{rs}. In that case, 
the confinement of gravity on the brane requires that bulk spacetime must be 
anti-deSitter (AdS) with negative $\Lambda$ in the bulk. It has been shown 
for the AdS bulk and flat brane system that the Newtonian gravity can be 
recovered on the brane with high energy $1/r^3$ correction to the potential 
\cite{rs,gt}. The most interesting case is of Schwarzschild - AdS bulk 
harbouring FRW brane. In that case localization of gravity on the FRW brane 
requires non-negative effective $\Lambda$ on the brane \cite{rk,pd}. \\

 We have established that universality of the force entirely determines its 
dynamics and it is nothing but the Einsteinian gravity \cite{n1}. We shall now 
argue that universality also determines its attractive character. Like 
electric charge for electric field, the charge for gravitational field is 
matter/energy distribution. It has only one polarity and calling it positive 
or negative is only a matter of convention. We do however know that stability 
of a system under any force demands charge neutrality. That is why all 
freely existing bodies like stars, planets down to atoms are all electrically  
charge neutral. So must also be the case for gravity. How could that be 
because it has only one kind of polarity? Charge neutrality has to be 
achieved for there exist large scale stable systems in the universe under 
gravity. Is there anything new which could be invoked to obtain neutrality? 
Yes, there is the gravitational field, which must now have gravitational 
charge of polarity opposite of the matter/energy distribution. The 
field must therefore have negative energy and that is why it must be 
attractive. It is again the universality that determines the attractive 
character of the field. In the Newtonian theory, it is an assumption that 
gravity is attractive which has been taken over in the Einsteinian theory. 
Attractive character is however associated with the spin 2 character of the 
field. Here we have articulated a very simple and novel argument based purely 
on the classical considerations. That the field is non-abelian and further 
following this line of argument it is also possible to work up an independent 
and new derivation and motivation for general relativity \cite{n4}. \\

 It is however well known that charge neutrality is a necessary condition for 
stable equilibrium but not sufficient. The Earnshaw's theorem states that it 
is impossible to attain stable equilibrium purely under 
electromagnetic force. This is because the field has two kinds of charges 
which are isolated and localized. On the other hand in the case of gravity 
the other (negative) charge is distributed all over the space and hence is 
not isolated and localized. A gravitational situation could be envisioned as 
follows: A positive charge (body) sitting in its own field which has negative 
charge spread around it in space like a net. This system is obviously stable. 
It is the distributed nature of the other charge that provides the stability. 
It is thus no surprise that systems bound by gravity are always stable. \\

 Let us for a moment digress to quantum theory. What question should we ask 
which can lead us from classical to quantum mechanics? We have two kinds of 
motion, particle and wave. Like particle wave must also carry energy and 
momentum with it. So it must like particle have a $4$-momentum vector while 
on the other hand its motion is completely determined by the $4$-wave vector. 
Since both these vectors refer to the same wave, they must be proportional. 
This gives the basic quantum mechanical relations between energy and 
frequency, and momentum and $3$-(wave)vector. From this it is easy to get to 
the uncertainty and commutation relations which form the basic quantum 
principle. \\

 The quantum principle is universal. Going by the guiding principle of 
universality, it must be expressible as a property, like the speed of light, 
of spacetime. This has unfortunately not happened. What is required is 
exactly what Minkowski did to SR by synthesizing the speed of light into 
the spacetime structure. This is however very difficult because synthesis of 
quantum principle with the spacetime would ask for discrete structure which 
is in contradiction with the inherent continuum of spacetime. However so long 
as this doesn't happen quantum theory will remain incomplete. Thus for 
completion of quantum theory it would be required that spacetime must have 
micro-structure which could accord to quantum principle. It is the geometry 
of that which would synthesize quantum principle with the spacetime. This 
is an open question of over 100 years standing. \\

 The same question is also coming up from the gravity side as well. Unlike 
quantum theory, GR is complete like classical electrodynamics. At high energy 
we have quantum electrodynamics. At high energy we know matter attains 
quantum character,i.e. $T_{ab}$ on the right becomes quantum. On the left is 
the spacetime which would now have to become quantum - discrete. This is what 
being pursued in the canonical approach of loop quantum gravity \cite{lqg}. 
In the string theory approach to quantum gravity, gravity is considered as 
a spin $2$ gauge field in higher dimensions against a fixed flat spacetime 
background and GR appears as a low energy effective theory \cite{gw}. The two 
approaches appear to be complimentary, either catching some aspect of the 
problem. Asymptotically the two would have to converge when the complete 
theory comes about. \\ 

 Both gravity at high energy as well as completion of quantum theory require 
discrete micro-structure for spacetime. This suggests that there may 
perhaps be one and the same answer to both the questions. In a sense the two 
approaches, string theory and loop quantum gravity anchor respectively to 
quantum and gravity. \\

 Adhering to our guiding principle of universality, what question should we 
ask and how should we enlarge the existing framework to answer the question? 
Since spacetime is the fundamental universal entity, let us ask do there 
still remain some properties of it which are yet untapped? One is 
dimensionality of space and the other is its (non) commutativity. The former 
is 
however essential for the string theory and is also quite in vogue in the 
brane world models while non commutativity of space has also attracted some 
attention 
recently and it is hoped that it might facilitate in building up discrete  
structure for space. \\

 Returning to the enigmatic $\Lambda$, we would like to argue that it is 
really anchored on the micro-structure of vacuum and hence might hold the 
ultimate key to the problem. It may in a deep sense connect through some 
duality relation micro with macro structure. It defines a new scale given 
by the Einsteinian gravity. On the other hand we already have the Planck 
length which is not given by any theory but constructed by using three 
universal constants. I believe that it is not a wise thing to take the Planck 
length a priori fundamental but instead we should try to deduce it in a 
fundamental manner. $\Lambda$ on the other hand is a scale provided by a 
fundamental theory and hence should be respected. It is natural to expect 
that there should exist a relation between them which will perhaps encode a 
profound physical truth. \\
 
 Let me come back to the guiding question one should ask? In 
the gravitational field equation, we have the curvature of spacetime on the 
left, and matter stress tensor and the vacuum energy $\Lambda$ on the right. 
If $T_{ab}$ lives only in $3$-space/brane which becomes quantum at high 
energy, while vacuum energy can still support a continuum bulk spacetime. 
Taking the cue from what we have discussed so far what question should we ask 
and how should we enlarge the framework to admit and answer the question? The 
universality demands that the Einstein equation should remain valid whether 
the matter field source is classical or quantum. For the quantum case, 
the spacetime curvature will be required to have discrete quantum character. 
Should that mean that spacetime itself becomes quantum? This is what is being 
explored by the canonical loop quantum gravity approach \cite{lqg}. It is 
then not an enlargment but rather adoption of a radically different framework.
 In this approach we remain bound to the $4$-dimensional spacetime and there 
is no background spacetime relative to which spacetime is being quantized. In 
the string theory approach, we are already in higher dimensions and gravity 
is being considered as massless spin 2 field and GR appears as the low energy 
effective theory in $4$ dimensions along with plethora of other fields 
\cite{gw}. In either case it is adoption of totally new framework which 
cannot be seen as enlargment of the existing framework. \\

 We believe and as we have argued above that gravity is a higher dimensional 
interaction. Perhaps impossibility of constructing a renormalizable theory of 
massless spin 2 particles in $4$ dimensions is indication of this fact.
The dynamics of the universal field should also determine the dimensionality 
of spacetime. One aspect of enlargment would therefore be lifting above 
$4$ dimensions. And spacetime curvature has to become discrete. One possible 
enlargment of the framework could be for matter fields confined to $3$-brane, 
and the $5$-dimensional bulk spacetime with $\Lambda$ support could provide 
the continuum background for gravity (spacetime curvature) to be quantized 
on the $3$-brane. This is the suggestion that crops up in the spirit of what 
has been done in going from Newton to Einstein. However the pertinent 
question is whether it is technically and conceptually workable? This needs 
to be investigated. \\

 Let me reiterate that we have here enunciated a method of asking question 
which is motivated by the principle of universality, and the question also 
suggests enlargment of the framework such that it gets answered. And we 
arrive at a new framework. One of the remarkably interesting applications of 
this method is to establish why the universal force has to be attractive? 
This is 
perhspas the simplest and most direct demonstration of why gravity is 
attractive. Following this train of thought what we need to do is to ask the 
right kind of question which will show us the way beyond GR or quantum 
theory. \\

 Finally, we have argued quite forcefully that $\Lambda$ cannot be zero and 
hence any quantization scheme should have to address to it \cite{gp}. I 
think that there must exist some basic link connecting $\Lambda$ with the 
Planck length which needs to be discovered. That may hold the key to the 
problem. Of course the problem is all encompassing and so involved that there 
may not be one key but we may have to integrate several of them together. \\

 These are some of the rumbling thoughts which I wanted to share and the more 
of them could be read in \cite{n4}.

\end{document}